\documentstyle[prl,aps,twocolumn,epsfig]{revtex}
\begin{document}
\draft
\title{Beyond the First Recurrence in Scar Phenomena}
\author{D. Wisniacki$^{1,2}$, F. Borondo$^1$, E. Vergini$^3$,
   and R. M. Benito$^4$}
\address{$^1$Departamento de Qu\'\i mica, C--IX,
   Universidad Aut\'onoma de Madrid,
   Cantoblanco--28049 Madrid, Spain.}
\address{$^2$Departamento de F\'\i sica ``J.J. Giambiagi'',
   FCEN, UBA, Pabell\'on 1, Ciudad Universitaria,
   1428 Buenos Aires, Argentina.}
\address{$^3$Dep. de F\'\i sica,
   Comisi\'on Nacional de Energ\'\i a At\'omica,
   Avenida del Libertador 8250,
   1429 Buenos Aires, Argentina.}
\address{$^4$Departamento de F\'\i sica y Mec\'anica,
   E.T.S.I. Agr\'onomos,
   Universidad Polit\'ecnica de Madrid,
   28040 Madrid, Spain.}
\date{Received August 12, 1999}
\maketitle

\begin{abstract}
The scarring effect of short unstable periodic orbits up to times of
the order of the first recurrence is well understood.
Much less is known, however, about what happens past this
short--time limit.
By considering the evolution of a dynamically averaged wave packet,
we show that the dynamics for longer times is controlled by only a few
related short periodic orbits and their interplay.
\end{abstract}
The study of the quantum manifestations of classical chaos
is at present a topic of very active research interest
\cite{Gutzwiller1}.
Great advance came from random matrix theory (RMT) which provides an
understanding of universal statistical properties of quantum spectra
\cite{Bohigas}.
The most striking departure from RMT described so far in the literature
is the phenomenon known as ``scar'' \cite{Kaplan99}.
This term describes an anomalous localization of quantum probability
density along unstable periodic orbits (PO) in classically chaotic
systems. Heller showed \cite{Heller84} the importance of the first
recurrences of POs in the scarring effect.
In a subsequent work Tomsovic and Heller \cite{TomsovicH}
demonstrated that the semiclassical propagation can be carried out
with remarkable precission long after classical fine structure had
developed on a scale much smaller than $\hbar$,
by computing the corresponding correlation function, $C_{\rm scl}(t)$,
as a sum of contributions of the homoclinic excursions of the PO.
This procedure is however cumbersome, since in general many orbits are
needed to obtain converged results, and this number increases rapidly
with time (see Ref.\ \onlinecite{TomsovicH} for details).
This picture is greatly simplified if alternatively considering the
corresponding averaged dynamics for finite periods of time.
In this case, as will be shown in this Letter, a structure of a few short
POs emerges that govern the quantum dynamics for times past the first
recurrence of the original PO.

Understanding scarring can be tackled from two sides.
One way is trying to disentangle the complexity involved in the
distribution of individual levels in the spectra of classically
chaotic systems \cite{Takami,Arranz,VerginiW}.
For example, in Ref.\ \cite{VerginiW} structures localized on short
POs of the stadium were obtained, by considering state correlation
diagrams and iteratively removing the parametric interactions
(avoided crossings) between the involved eigenstates.
The other way consists of approaching the problem in a much more
straighforward fashion, by studying how the dynamics of POs
induce scars in the eigenfunctions of the system.

Heller's work provided a time--dependent view that shows
how recurrences in the short time dynamics of a wave
packet along the neighborhood of an isolated PO produces the
accumulation of quantum probability density characteristic of scars.
Very recently Kaplan and Heller\cite{KaplanX} have shown how the use
of coherent wave packet sums, decaying as the log--time, leads to
enhanced scarring.
Later, it was described by some of us \cite{Polavieja} how
(nonstationary) wave functions highly localized over POs can be
constructed from finite time Fourier transform of wave packets.
This method allows the calculation of the contribution of each
eigenstate to scars.

The time--dependent approach presents the additional advantage
of being easily connected with spectroscopy experiments through
Fourier transform into the energy domain of the autocorrelation
function generated by a test initial wave packet \cite{Heller91}.
In the short times limit, recurrences originated by isolated unstable
POs in the vicinity of the packet translate into low resolution
features of the corresponding spectra (or local density of states)
consisting of peaks or bands whose widths are determined by their
Lyapunov exponents.
In the other limit, corresponding to long times, non--linear effects
introduce modulations in these envelopes leading essentially to
Porter--Thomas distributions of spectral intensities, although
it has been shown recently \cite{Kaplan} that the scarring effect
is still noticeable in the tail of the distribution.
Finally, when the Heisenberg time is reached, individual eigenstates
are resolved, and we have essentially the infinite resolution version
of the spectrum.
In this sense, the overall aspect of chaotic spectra is highly
conditioned by the least unstable PO (LUPO) contained in the phase
space region spanned by the test wave packet.

In this Letter we consider the case of intermediate times
which allow the participation of orbits longer than the LUPO.
The associated low resolution spectra split into several components
showing evidence of the participation of only a few localized structures
corresponding to short POs, which also present their signatures in the
correlation function.
It is the magnitude of the interaction (in the sense of
Ref.\ \onlinecite{VerginiW}) among these localized wave functions and
that associated to the LUPO that determines which orbits come into play.

In our study we used the desymmetrized stadium billiard of radius
$r=1$ and enclosed area of $1+\pi/4$.
This system constitutes a paradigm in the study of hard chaos.
From the classical point of view it has been demonstrated to be ergodic.

To study the scarring effect of the different stadium POs we closely
follow the method described in Ref.~\cite{Polavieja}, which consists
in following for finite times the dynamics of a wave packet initially
located in the vicinity of a particular PO.
In our case we use a harmonic oscillator coherent state (throughout
the paper $\hbar$ is set equal to one),
\begin{equation}
   \langle x,y |\phi \rangle =
   \left[ \frac{2\alpha}{\pi} \right]^{1/4}
   {\rm e}^{- \alpha(x-x_0)^2 - \alpha(y-y_0)^2}  \;
   {\rm e}^{i (P_x^0 x + P_y^0 y)},
 \label{eq:phi0}
\end{equation}
with $\alpha$=30.68.
The time evolution of this state can be followed by projection onto the
stadium eigenstates, $|n\rangle$:
\begin{equation}
    |\phi(t)\rangle = e^{-i \hat{H} t} |\phi(0)\rangle =
      \sum_n | n \rangle \langle n | \phi(0) \rangle \; e^{-i E_n t},
 \label{eq:phit}
\end{equation}
which are calculated by the scaling method \cite{VerginiS}.
Recurrences in the corresponding correlation function,
$C(t)=\langle \phi(0) | \phi(t) \rangle$, determine the structure
in the associated finite resolution spectrum,
\begin{equation}
   I_T (E) = \frac{1}{2 \pi} \int_{-\infty}^\infty \; dt \;
     C(t) \; W_T(t) \; e^{i E t}
 \label{eq:I_T}
\end{equation}
where a window function, $W_T(t)$, has been introduced to eliminate
features in $C(t)$ taking place after a given time.
Choosing a Gaussian form, $e^{-t^2/2T^2}$, for this window the
following expression for $I_T(E)$ is obtained
\begin{equation}
   I_T (E) = \frac{T}{(2 \pi)^{1/2}} \sum_n
   | \langle n | \phi(0) \rangle |^2 \; e^{-T^2 (E-E_n)^2/2}.
 \label{eq:I_Texp}
\end{equation}
As shown in Ref.~\cite{Polavieja}, it is also possible to compute the
wave functions associated to each low resolution band (or envelope),
by means of the expression
\begin{eqnarray}
   | \Psi^{E_0} \rangle & = &
     \frac{1}{2 \pi} \int_{-\infty}^\infty \; dt \;
     | \phi(t) \rangle W_T(t) \; e^{i E_0 t} \nonumber \\
     & = & \frac{T}{(2 \pi)^{1/2}} \sum_n
     | n \rangle \langle n | \phi(0) \rangle \; e^{- T^2 (E_0-E_n)^2/2}.
 \label{eq:wfband}
\end{eqnarray}
These band wave functions have been shown to be highly localized
along the scarring PO, when the time of the first recurrence
is used as the smoothing time \cite{Polavieja}.

\begin{figure}
\epsfig{file=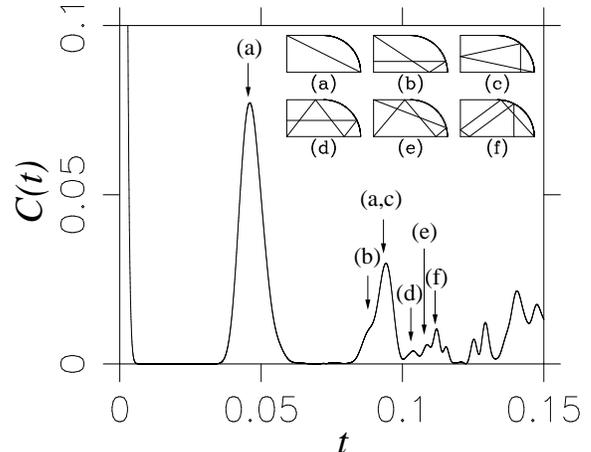,width=6cm,angle=-90}
\caption{Autocorrelation function corresponding to a wave packet
  initially centered on the diagonal periodic orbit (a) of a
  desymmetrized stadium billiard with $r=1$ and area $1+\pi/4$.
  This and other orbits relevant to our study are shown in the inset.
  Arrows indicate the recurrence times of these periodic orbits.}
\label{fig:C(t)POs}
\end{figure}

We will concentrate our study in one of the most representative short
PO of the desymmetrized stadium;
namely, that running along the diagonal joining the two extreme points
on both axis, which corresponds to a diamond shape in the full version
of the stadium.
This PO is presented in the inset to Fig.~\ref{fig:C(t)POs} along with
some others which are relevant to our study.

\begin{figure}
\epsfig{file=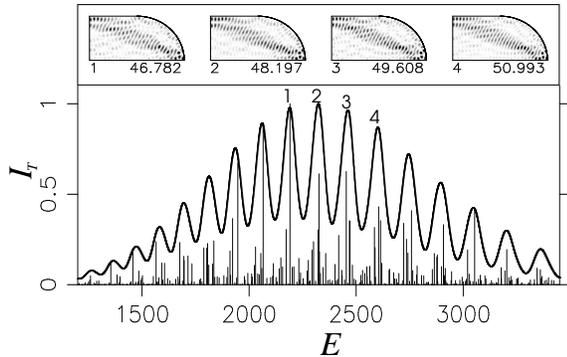,scale=0.4,angle=-90}
 \caption{Infinite resolution spectrum (sticks) and smoothed version
  of it for $T=0.04$ (full line) corresponding to the autocorrelation
  function of Fig.~\protect\ref{fig:C(t)POs}.
  In the upper part, the wave functions associated to bands 9 to 12,
  calculated using Eq.~(\protect\ref{eq:wfband}) and $T=0.04$,
  are presented.
  The corresponding values of $\protect\sqrt{\langle E \rangle}$ are
  shown below each plot.}
 \label{fig:I_Twf1}
\end{figure}

Using a wave packet [Eq.~(\ref{eq:phi0})] initially located on the
middle of the diagonal PO
[$(x_0,y_0,P_x^0,P_y^0)=(1,1/2,96/\sqrt{5},-48/\sqrt{5})$],
the autocorrelation function and the corresponding infinite ($T=\infty$)
and low resolution ($T=0.04$) spectra have been computed.
The results are presented in Figs.~\ref{fig:C(t)POs} and
\ref{fig:I_Twf1}, respectively.
Notice that the time used in the smoothed version of the spectrum is
roughly equal to the period of appearance of the first recurrence in
$C(t)$, to which only the diagonal PO contributes.
It can be observed that the peaks corresponding to $I_\infty$ appear
grouped in clumps, so that $I_T$ exhibit a series of (seventeen) very
well defined bands, which are equally spaced.
Moreover, the positions of these peaks, calculated as the mean energy
value of each band, are in very good agreement with the energies
obtained from the wave numbers, $k$, quantized according to the usual
Bohr--Sommerfeld quantization rule:
\begin{equation}
   k = \frac{2 \pi}{L} \left( n + \frac{\nu}{4} \right)
 \label{eq:BS}
\end{equation}
where $n$ is the number of nodes along the orbit, $L = 2 \sqrt{5}$ its
lenght, and $\nu = 9$ the corresponding Maslov index.
As an illustration, we show in the upper part of Fig.~\ref{fig:I_Twf1}
the wave functions (squared) associated to bands 9 to 12,
calculated using Eq.~(\ref{eq:wfband}).
The values of $\sqrt{\langle E \rangle}$ are also given below each plot,
so that the agreement with the value predicted by the quantization
condition (\ref{eq:BS}) can be checked.
As expected, the wave functions appear highly localized along the
diagonal PO, corresponding to a number of nodes of 31 to 34,
respectively.
This result could have also been obtained with the more elaborated
coherent wave packets sums of Kaplan and Heller \cite{KaplanX}.

Let us consider now recurrences taking place at longer times
in the correlation function of Fig.~\ref{fig:C(t)POs}.
One can forsee that their analysis will be more complicated;
for one thing they appear at longer times so that the packet has had
the opportunity to explore more extended regions of phase space
(where the linearized dynamics around the PO is no longer valid),
and also several orbits contribute to them.
The mechanism responsible for the dynamical coupling of these POs,
accessed by the phase space sampling associated to the choice of the
initial wave packet (\ref{eq:phi0}), is not obvious.
Some preliminary results \cite{inprep} indicate that the dynamics along
the manifolds emanating from them and their proximity, is an important
factor in this issue.

\begin{figure}
\epsfig{file=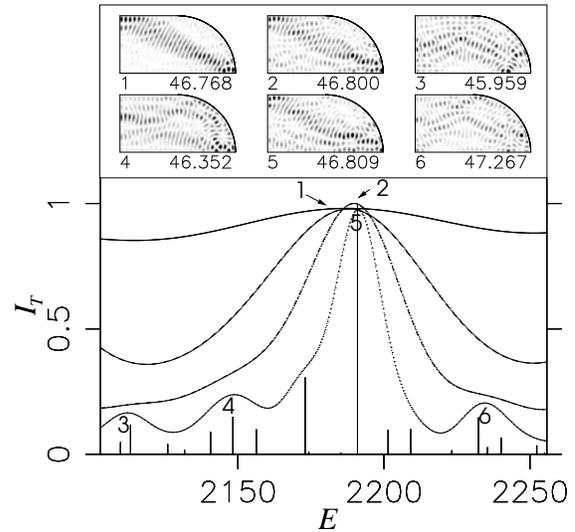,width=7cm,,angle=-90}
 \caption{Enlargement of the region of the spectrum of
  Fig.~\protect\ref{fig:I_Twf1} corresponding to the 9--th band.
  The smoothed version has been computed for three different values
  of the resolution time, $T$: 0.02 (full line), 0.04 (dashed line),
  0.07 (dot--dashed line) and 0.135 (dotted line).
  In the upper part, the wave functions associated to the labeled
  bands are presented.
  The corresponding values of $\protect\sqrt{\langle E \rangle}$ are
  shown below each plot.}
 \label{fig:I_Twf2}
\end{figure}

For this purpose, let us examine the group of peaks in $C(t)$ at
$t \sim 0.08-0.12$.
It presents a complicated structure; in addition to the main
maximum at $t \sim 0.09$, it exhibits a shoulder at smaller time
values and four much shorter peaks at longer times.
Taking into account the periods of the orbits presented in
Fig.~\ref{fig:C(t)POs} a tentative assignment of the contribution
to each peak has been made.
The result is indicated by labeled arrows over the curve of $C(t)$.
To check our hypothesis, we consider a series of low resolution
spectra computed using progressively larger smoothing times, so that
POs with increasingly longer periods are allowed to come into play.
Some representative results, corresponding to the 9--th band of
the spectrum of Fig.~\ref{fig:I_Twf1}, are presented in
Fig.~\ref{fig:I_Twf2}.
Our calculations show that in the range $T=0.02-0.07$ the smoothed
band consists of only one peak, which begins to develop an incipient
intraband structure as $T$ increases (see different curves in
Fig.~\ref{fig:I_Twf2}).
After that time the structure gets more clearly defined, so that for
the interval $T=0.12-0.20$ four intraband components are readily
observable (dotted line in Fig.~\ref{fig:I_Twf2}).
The associated wave functions are shown in the upper part of the
figure.
As can be seen, the wave function labeled by 2, corresponding to
$T=0.07$, is very similar to that already presented in
Fig.~\ref{fig:I_Twf1} that was computed at $T=0.04$.
However, function 1, calculated at a smaller time, presents a much
stronger localization along the diagonal PO, being better defined
specially at the corners.
This is due to the fact that a smaller resolution time allows the
participation of more eigenstates in the expansion of the wave
function associated to each band.
Finally, at the largest time considered, $T=0.135$, the band splits
into four components, whose associated wave functions are shown in
plots 3 to 6 of Fig.~\ref{fig:I_Twf2}.
By visual inspection, these functions are found to be scarred by POs
(d), (c), (b)/(a) and (d), respectively.
This assignment was confirmed more quantitatively in the following way.
In the first place, we constructed localized wave functions on the
scarring orbits in the same way described above for the diagonal PO
(see Fig.~\ref{fig:I_Twf1}), and calculated the overlaps of the
resulting functions with those of the intrabands considered.
The wave function associated to peak number 5 is a linear combination
(56\% and 42\%) of the structures localized on orbits
(b) and (a), being the corresponding overlap values 0.75 and 0.65,
respectively. Values greater than 0.7 were obtained for the other three
peaks. In the second place,
the corresponding calculated values of $\sqrt{\langle E \rangle}$:
45.959, 46.352, 46.809, and 47.267,
agree remarkably well with the predicted values of $k$ that are
obtained from the Bohr--Sommerfeld quantization rule
[Eq.~(\ref{eq:BS}) with the appropriate values for the parameters
$L$ and $\nu$]: 45.966, 46.367, 46.748/46.715, and 47.225.
The reason why the 9--th band in the spectrum of
Fig.\ \ref{fig:I_Twf1} splits into these components can be
understood by considering the Hamiltonian matrix elements among
localized states 1 and 3--6 of Fig.\ \ref{fig:I_Twf2}.
These numbers, defined as interactions in Ref.\ \onlinecite{VerginiW},
when calculated (226, 664, 1706 and 533 respectively)
turn out to be larger than any other interaction
with localized structures in this energy region.
The picture that emerges is analogous to that of non--overlapping
resonances embedded in a continuum (see discussion in
Ref.\ \onlinecite{Llorente}).
The band or intraband structures obtained with Eq.\ (\ref{eq:wfband})
correspond, loosely speaking, to states dynamically averaged over
classical paths. They are initially located in the neighbourhood of
unstable POs (equivalent to a decaying resonance) which are the only
renmant of order in the middle of the surrounding chaotic sea (playing
the role of the continuum).
Moreover, the corresponding spectra present a great resemblance with
resonance spectra, and the rhs of Eq.\ (\ref{eq:wfband}) can be viewed
as an approximate projection operator (on each resonance) acting on
$|\phi(0)\rangle$.

\begin{figure}
\epsfig{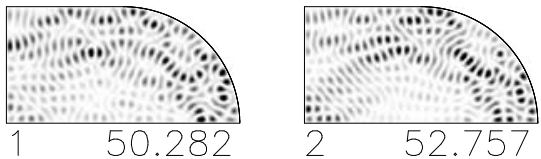}
\vspace{0.5cm}
 \caption{Wave functions corresponding to different components of
  bands 12 and 13 in the spectra of Fig.~\protect\ref{fig:I_Twf1},
  showing localization along periodic orbits (e) and (f) of
  Fig.~\protect\ref{fig:C(t)POs}.
  The corresponding values of $\protect\sqrt{\langle E \rangle}$ are
  shown below each plot.}
\label{fig:wfband3}
\end{figure}

The same analysis can be applied to investigate the intraband
structure of other bands in the spectrum of Fig.~\ref{fig:I_Twf1},
in order to discover the signature of other POs.
This is the case, for example, of the wave functions presented in
Fig.~\ref{fig:wfband3}, which are scarred by POs (e) and (f),
respectively.
They correspond to peaks in the intraband structure of bands 12 and 13
of Fig.~\ref{fig:I_Twf1}.
Again the computed values of $\sqrt{\langle E \rangle}$:
50.282 and 52.757, agree very well with the predicted values of $k$:
50.245 and 52.722.

This result indicates that the intraband structure associated to
different bands can be quite distinct.
As shown, this is due to the fact that all POs involved in a given
intraband pattern (corresponding to a given smoothing time) quantize in
different ways, so that each one gives rise to a characteristic spacing.
It is the relation among the values of these spacings and that of the
original orbit (the LUPO in our case), which determines the distribution
of new peaks in each band.
Since they are in principle unconnected, the pattern under each band
will be different.

In conclusion, we have presented a study of the scarring effect of
POs beyond the short time dynamics in a very chaotic system.
A few POs, other than the LUPO, have been shown to be able to
develop their influence in the low resolution features of the
corresponding spectra or local density of states, appearing at the
associated quantized energy values; in other words they are responsible
for the averaged dynamics in this time range.
It has also been  described how localized wave functions can be
constructed for this intermediate time domain.
These results provide a good smoothed picture around a given PO and
the family of orbits connected to it,
and then constitute an important first step in order to fully
disentangle the complexity involved in the eigenvalues spectrum
of very chaotic systems.

This work has been supported in part by DGES (Spain) under Projects
No.~PB95--425 and PB96--76.
DW gratefully acknowledges support from Agencia Espa\~nola de
Cooperaci\'on Internacional (Spain) and FOMEC (Argentina) for
his stay in Madrid.
%

\end{document}